\documentclass[a4paper,12pt]{article}
\usepackage{amsfonts,slashed}
\usepackage{url}
\usepackage{latexsym}
\usepackage{amsmath}
\usepackage{amssymb}
\usepackage{indentfirst}
\usepackage{graphicx}
\usepackage{epsfig}
\usepackage{amsthm}

\usepackage{ifpdf}
\ifx\pdfoutput\undefined
   \pdffalse
   \usepackage{cite}
 \else
   \pdfoutput=1
   \pdftrue
  \usepackage[pdftex]{hyperref}
\fi

\numberwithin{equation}{section}
\def\be{\begin{equation}}
\def\ee{\end{equation}}
\def\ba{\begin{array}}
\def\ea{\end{array}}

\newcommand{\bea}{\begin{eqnarray}}
\newcommand{\eea}{\end{eqnarray}}

\def\ii{{\rm i}}

\newcommand{\bbox}{\lower.2ex\hbox{$\Box$}}

\def\bfone{\relax{\rm 1\kern-.35em 1}}
\def\bfzero{\relax{\rm I\kern-.18em 0}}

\begin{document}

\title{\bf{Hidden Role of Maxwell Superalgebras \\ in the Free Differential Algebras \\ of $D=4$ and $D=11$ Supergravity}}
\author{{\bf{Lucrezia Ravera}}$^{\star}$\thanks{lucrezia.ravera@mi.infn.it} \\ \\
\vspace{2cm} 
{\small $^{\star}$\textit{INFN, Sezione di Milano, Via Celoria 16, I-20133 Milano, Italy}}}
\maketitle

\begin{abstract}

The purpose of this paper is to show that the so-called Maxwell superalgebra in four dimensions, which naturally involves the presence of a nilpotent fermionic generator, can be interpreted as a hidden superalgebra underlying $\mathcal{N}=1$, D=4 supergravity extended to include a 2-form gauge potential associated to a 2-index antisymmetric tensor. In this scenario, the theory is appropriately discussed in the context of Free Differential Algebras (an extension of the Maurer-Cartan equations to involve higher-degree differential forms). 
The study is then extended to the Free Differential Algebra describing D=11 supergravity, showing that, also in this case, there exists a super-Maxwell algebra underlying the theory. 

The same extra spinors dual to the nilpotent fermionic generators whose presence is crucial for writing a supersymmetric extension of the Maxwell algebras, both in the D=4 and in the D=11 case, turn out to be fundamental ingredients also to reproduce the D=4 and D=11 Free Differential Algebras on ordinary superspace, whose basis is given by the supervielbein. The analysis of the gauge structure of the supersymmetric Free Differential Algebras is carried on taking into account the gauge transformations from the hidden supergroup-manifold associated with the Maxwell superalgebras.

\end{abstract}

\newpage

\section{Introduction}

It is well known that supergravity theories in $D \geq 4$ space-time dimensions contain gauge potentials described by $p$-forms, of various $p > 1$, associated to $p$-index antisymmetric tensors. In this scenario, the Free Differential Algebras framework, that is an extension of the Maurer-Cartan equations to involve higher-degree differential forms, is particularly well suited for studying supergravity models. The concept of Free Differential Algebra (FDA in the sequel) was introduced in \cite{Sullivan} and subsequently applied to the study of supergravity theories (see, for instance, Ref. \cite{D'AuriaFre}). 

A review of the standard procedure for the construction of a minimal FDA (namely a FDA where the differential of any $p$-form does not contain forms of degree greater than $p$) starting from an ordinary Lie algebra can be found in \cite{Hidden}.

In \cite{D'AuriaFre}, the authors considered the $D=11$ supergravity theory of \cite{Cremmer}, introducing and investigating the supersymmetric FDA describing the theory (using the so-called superspace geometric approach) in order to see whether the FDA formulation could be interpreted in terms of an ordinary Lie superalgebra (in its dual Maurer-Cartan formulation). This was proven to be true, and the existence of a hidden superalgebra underlying the $D=11$ supergravity theory was presented for the first time. It includes the $D=11$ Poincar\'{e} superalgebra as a subalgebra, but it also contains two extra, almost-central, bosonic generators, which were lately understood as $p$-brane charges, sources of the dual potentials $A^{(3)}$ and $B^{(6)}$ appearing in the (complete) FDA of \cite{D'AuriaFre} (see Refs. \cite{Hull:1994ys}, \cite{Townsend:1995gp}).

Furthermore, a nilpotent fermionic generator must be included to close the superalgebra and in order for the same superalgebra to reproduce the $D=11$ FDA on ordinary superspace, whose basis is given by the supervielbein. Relevant contributions concerning the physical role played by this extra fermionic generator were given first in \cite{vanHolten:1982mx} and then in particular in \cite{Bandos:2004xw}, \cite{Bandos:2004ym}, where the results presented in \cite{D'AuriaFre} were further analyzed and generalized. Finally, its group-theoretical and physical meaning was recently clarified in \cite{Hidden} (and subsequently further discussed in \cite{Malg}): In \cite{Hidden} it was shown that the spinor $1$-form dual to the nilpotent fermionic charge is not a physical field in superspace, rather behaving as a cohomological BRST ghost, since its supersymmetry and gauge transformations exactly cancel the non-physical contributions coming from the extra tensor fields, guaranteeing that the extra bosonic $1$-forms dual to the almost-central charges are genuine abelian gauge fields.\footnote{Actually, as it was lately pointed out in \cite{Malg}, the extra spinor $1$-form dual to the nilpotent fermionic generator can be parted into two different spinors, whose integrability conditions close separately.}

As shown in Ref. \cite{Hidden}, where the authors analyzed also the FDA of the minimal $\mathcal{N} = 2$, $D = 7$ supergravity theory, this interpretation is valid for any supergravity theory containing antisymmetric tensor fields, and any supersymmetric FDA can always be traded for a hidden Lie superalgebra
containing fermionic nilpotent generators (see also \cite{SM4} for the study of a particular $D=4$ FDA case).

In the first part of this paper, we will consider the FDA of $\mathcal{N}=1$, $D=4$ supergravity containing a $2$-form potential under the same perspective of \cite{Hidden}. Let us mention, here, that supergravity in $D=4$ space-time dimensions is often formulated as a theory of gravity coupled to scalar-vector multiplets only, that is to say $1$-form gauge fields. On the other hand, when we think of the theory as obtained by Kaluza-Klein compactification from eleven-dimensional supergravity, then it naturally contains also $2$-form fields (tensor multiplets). In four dimensions, if these $2$-form fields are massless, then they can be dualized, through Hodge duality of their field strengths, to scalars (this is the reason why they often do not explicitly appear in the formulation). However, when they are massive\footnote{This happens, for instance, in the case in which the higher-dimensional theory is reduced via a flux compactification.} such dualization does not (at least directly) apply and, in this case, the $2$-form gauge fields must be made manifest \cite{Gunaydin:2005df} (see also Refs. \cite{Andrianopoli:2007ep}, \cite{Andrianopoli:2011zj} and \cite{sito} for more details on the role of $2$-forms in four-dimensional supergravity theories).

The aim of the present paper is to show that the so-called \textit{minimal Maxwell superalgebra} (or \textit{minimal super-Maxwell algebra}) in four dimensions (a non-semisimple superalgebra naturally endowed with a nilpotent fermionic generator), can be interpreted as a hidden superalgebra underlying the FDA of $D=4$ supergravity that includes a $2$-form potential $A^{(2)}$. This will be done by studying the parametrization of $A^{(2)}$ and the hidden gauge structure of the FDA on the same lines of what was done in the $D=11$ (and $D=7$) case in \cite{D'AuriaFre}, \cite{Hidden}. Then, we will extend our discussion to the FDA introduced in \cite{D'AuriaFre}, which describes $D=11$ supergravity, showing that, also in this case, there exists a Maxwell superalgebra underlying the theory.\footnote{Actually, we will consider the $D=11$ FDA just containing a $3$-form potential $A^{(3)}$. We leave the study of the complete FDA containing also a $6$-form potential $B^{(6)}$ (see Refs. \cite{D'AuriaFre}, \cite{Hidden}) to future works.}
The extra spinors dual to the nilpotent fermionic generators whose presence is crucial for writing a supersymmetric extension of the Maxwell algebras, both in the $D=4$ and in the $D=11$ case, will turn out to be fundamental also to reproduce the $D=4$ and $D=11$ FDAs on ordinary superspace. 

This work is organized as follows: In Section \ref{Maxalg}, we first recall the main features of the Maxwell superalgebra; then, we move to the analysis of the hidden gauge structure of the supersymmetric FDA of $\mathcal{N}=1$, $D=4$ supergravity (containing a $2$-form potential $A^{(2)}$), showing that the Maxwell superalgebra can be viewed as a hidden superalgebra underlying the theory. Subsequently, in Section \ref{11d}, we extend our study and results to the FDA describing $D=11$ supergravity (which, in its minimal cohomology formulation, contains just a $3$-form potential $A^{(3)}$), introducing a (hidden) Maxwell superalgebra underlying the theory. Finally, Section \ref{Concl} contains the conclusions and possible future developments. In the Appendix we collect our conventions and some useful formulas.

\section{Minimal super-Maxwell algebra and hidden gauge structure of the D=4 supergravity FDA} \label{Maxalg}

After the discovery of the cosmic microwave background and the mysterious dark energy, it appears interesting to consider some field densities uniformly filling space-time. One such modification of empty Minkowski space can be obtained by adding a
constant electromagnetic field background, parametrized by additional degrees of freedom related to tensorial almost-central charges. The presence of a constant electromagnetic field modifies the Poincar\'{e} symmetries into
the so-called Maxwell symmetries. On the other hand, since the advent of supersymmetry, there has been a great interest in superalgebras going beyond the super-Poincar\'{e} one. 

In particular, the (minimal) Maxwell superalgebras are (minimal) super-extensions of the Maxwell algebra, which in turn is a non-central extension of the Poincar\'{e} algebra involving an extra, bosonic, abelian generator (along the lines of non-commutative geometry). 

Specifically, the $D=4$ Maxwell algebra is obtained by replacing the
commutator $[P_a, P_b] = 0$ ($a=0,1,2,3$) of the Poincar\'{e} algebra with $[P_a,P_b]=Z_{ab}$, where $Z_{ab}=-Z_{ba}$ are abelian generators commuting with translations and behaving like a tensor
with respect to Lorentz transformations (\textit{i.e.} $Z_{ab}$ are tensorial central charges).
Setting $Z_{ab}=0$ one gets back to the Poincar\'{e} algebra.

The Maxwell algebra arises when one considers symmetries of systems evolving in flat Minkowski space filled in by a constant electromagnetic background \cite{bacry}, \cite{schrader}. Indeed, an action for a massive particle which is invariant under the Maxwell symmetries satisfies the equations of motion of a charged particle interacting with a constant electromagnetic field via the Lorentz force. In particular, in order to interpret the Maxwell algebra and the corresponding Maxwell group, a Maxwell group-invariant particle model was studied on an extended space-time with coordinates $(x^\mu, \phi^{\mu \nu})$, where the translations of $\phi^{\mu \nu}$ are generated by $Z_{\mu \nu}$ \cite{gomis0}, \cite{gomis1}, \cite{Bonanos:2008ez}, \cite{gibbons}. The interaction term described by a Maxwell-invariant $1$-form introduces new tensor degrees of freedom, momenta conjugate to $\phi^{\mu \nu}$, and, in the equations of motion, they play the role of a background electromagnetic field which is constant on-shell and leads to a closed, Maxwell-invariant $2$-form.
The Maxwell algebra describes, at same time, the particle and the
constant electromagnetic background in which it moves.

Furthermore, in \cite{deAzcarraga:2010sw}, driven by the fact that it is often thought that the cosmological constant problem
may require an alternative approach to gravity, the authors presented a geometric framework based on the $D=4$ gauged Maxwell algebra, involving six new gauge fields associated with their abelian generators, and described its application as source of an additional contribution to the cosmological term in Einstein gravity, namely as a generalization of the cosmological term.
Subsequently, in \cite{Durka:2011nf} the authors deformed the $AdS$ algebra by adding extra non-abelian $Z_{ab}$ generators, forming, in this way, the negative cosmological constant counterpart of the Maxwell algebra. Then, they gauged this algebra and constructed a dynamical model. In the resulting theory, the gauge fields associated with the Maxwell-like generators $Z_{ab}$ appear only in topological terms that do not influence the dynamical field equations.

The minimal supersymmetric extension of the $D=4$ Maxwell algebra was obtained in \cite{gomis2} as a minimal enlargement of the $\mathcal{N}=1$ Poincar\'{e} superalgebra, by adding two four-dimensional Majorana supercharges ($Q_\alpha$ and $\Sigma_\alpha$, $\alpha=1,2,3,4$), and, mathematically optional, two scalar generators ($B_5$ and $B$). Thus, in terms of dual Maurer-Cartan $1$-forms, this minimal supersymmetrization of the Maxwell algebra naturally requires to introduce, besides the $1$-form spinor field $\psi^\alpha$ (dual to the supercharge $Q_\alpha$), also an extra Majorana $1$-form spinor field $\xi^\alpha$ (dual to the nilpotent fermionic generator $\Sigma_\alpha$).
The minimal Maxwell superalgebra introduced in \cite{gomis2} (and, subsequently, further discussed and deformed in \cite{gomis3}) seems to be specially appealing, since the coset $\frac{\text{super-Maxwell}}{\text{Lorentz} \times B_5}$ describes the supersymmetries of flat (Wess-Zumino) Minkowski superspace with arbitrary constant values of an abelian supersymmetric field-strength background.
In this set up, the superspace coordinates $(x^\mu, \theta^\alpha, \phi)$ are supplemented in the framework of Maxwell supergeometry by graded additional coordinates related to the generators $(\Sigma_\alpha, Z_{\mu \nu}, B)$.

At a later time, in \cite{Kamimura:2011mq} the authors wrote supersymmetrization schemes of the $D=4$ Maxwell algebra, and further generalizations of Maxwell (super)algebras where then derived and studied in the context of expansion of Lie (super)algebras \cite{deAzcarraga:2012zv}. Subsequently, the Maxwell superalgebra of \cite{gomis3} and its generalizations have been obtained through a particular expansion procedure that goes under the name of $S$-expansion, starting from the $AdS$ superalgebra \cite{Concha2}. This family of superalgebras, containing the Maxwell algebras type as bosonic subalgebras, can be viewed as a generalization of the D'Auria-Fr\'{e} superalgebra introduced in \cite{D'AuriaFre} and of the Green algebra \cite{green}.\footnote{The Green algebra was used in \cite{Siegel:1994xr} to produce a superstring action with a manifestly supersymmetric Wess-Zumino term. The procedure was further generalized in \cite{Bergshoeff:1995hm} to super $p$-branes by introducing larger Green-type superalgebras (see also \cite{Sezgin:1996cj}).}

Lately, in \cite{deAzcarraga:2014jpa} it was shown that the first-order $\mathcal{N}= 1$, $D=4$ pure supergravity Lagrangian $4$-form can be obtained geometrically as a quadratic expression in the curvatures of the Maxwell superalgebra.
Furthermore, in \cite{CR2} the authors presented the construction of the $D = 4$ pure supergravity action (plus boundary terms) starting from a minimal Maxwell superalgebra (which can be derived from $\mathfrak{osp}(4 \vert 1)$ by applying the $S$-expansion procedure), showing, in particular, that the $\mathcal{N} = 1$, $D = 4$ pure supergravity theory can be alternatively obtained as the MacDowell-Mansouri like action built from the curvatures of this minimal Maxwell superalgebra.
Remarkably, also in this context, the Maxwell-like fields do not contribute to the dynamics of the theory, appearing only in the boundary terms. Moreover, recently, in \cite{Concha:2016hbt} the authors introduced an alternative way of closing Maxwell-like algebras. 

For all the reasons listed above, the (super-)Maxwell algebras result to be very attractive in the context of (super)gravity theories. Let us now go deep in some technical details concerning the minimal $D=4$ Maxwell superalgebra of Ref. \cite{deAzcarraga:2014jpa}.

As we have already mentioned, besides the Poincar\'{e} generators, the minimal $D=4$ Maxwell algebra contains six additional tensorial charges $Z_{ab}$ that centrally extend the abelian translation algebra and behave tensorially under the Lorentz algebra. 

Then, the minimal $D=4$ super-Maxwell algebra is generated by $\lbrace J_{ab}, P_a, Z_{ab}, Q_\alpha , \Sigma_\alpha \rbrace$ ($a=0,1,2,3$, $\alpha=1,2,3,4$), and its (anti)commutation relations read:
\begin{align}
& [J_{ab} , J_{cd}] = \eta_{bc} J_{ad} - \eta_{ac} J_{bd}-\eta_{bd}J_{ac}+\eta_{ad}J_{bc}, \nonumber \\
& [J_{ab},Z_{cd}] =  \eta_{bc} Z_{ad} - \eta_{ac} Z_{bd}-\eta_{bd}Z_{ac}+\eta_{ad}Z_{bc}, \nonumber \\
& [J_{ab},P_c] = \eta_{bc}P_a - \eta_{ac}P_b, \nonumber \\
& [J_{ab}, Q_\alpha]= - \frac{1}{2}(\gamma_{ab} Q)_\alpha , \nonumber \\
& [J_{ab}, \Sigma_\alpha ]= - \frac{1}{2}(\gamma_{ab}\Sigma)_\alpha , \nonumber  \\
& [P_a,P_b]=Z_{ab}, \quad [P_a,Z_{cd}]=0, \nonumber \\
& [P_a , Q_\alpha] = - \frac{1}{2}(\gamma_a \Sigma)_\alpha , \nonumber \\
& [P_a,\Sigma_\alpha]=0,  \nonumber \\
& [Z_{ab},Z_{cd}]=0, \nonumber \\
& [Z_{ab},Q_\alpha]=0, \quad [Z_{ab},\Sigma_\alpha]=0, \nonumber \\
& \lbrace Q_\alpha , Q_\beta \rbrace = (\gamma^a C)_{\alpha \beta} P_a , \nonumber \\
& \lbrace Q_\alpha , \Sigma_\beta \rbrace = - \frac{1}{2}(\gamma^{ab} C)_{\alpha \beta} Z_{ab}, \nonumber \\
& \lbrace \Sigma_\alpha , \Sigma_\beta \rbrace = 0 , \label{smlast}
\end{align}
where $C$ stands for the charge conjugation matrix, $\eta_{ab}$ is the (mostly plus) Minkowski metric, $\gamma_a$ and $\gamma_{ab}$ are Dirac gamma matrices in four dimensions, $Q_\alpha$ is the supersymmetry generator, and where we can see that the $[P_a,Q_\alpha]$ commutator produces an extra, nilpotent, fermionic generator, $\Sigma_\alpha$; in particular, the latter naturally appears in the supersymmetric extension of the Maxwell algebra. 

Let us notice that the Lorentz-type algebra generated by $\lbrace J_{ab},Z_{ab} \rbrace$ is a subalgebra of the above superalgebra. This Maxwell superalgebra can also be obtained by imposing $\tilde{Z}_{ab} = 0$ in the generalized minimal super-Maxwell algebra of \cite{Concha2}, \cite{CR2}, which in turn can be derived from the $\mathfrak{osp} (4 \vert 1)$ superalgebra by applying the abelian semigroup expansion procedure ($S$-expansion, for short). 

We shall describe the superalgebra given in (\ref{smlast}) through its Maurer-Cartan equations satisfied by the set of $1$-form fields $\sigma^A= \lbrace \omega^{ab}$, $V^a$, $B^{ab}, \psi^\alpha, \xi^\alpha \rbrace$ dual to the set of generators $T_A = \lbrace J_{ab}, P_a, Z_{ab}, Q_\alpha , \Sigma_\alpha \rbrace$ (in the sequel, we will neglect the spinor index $\alpha$, for simplicity), that is to say
\begin{align}
& \omega^{ab} (J_{cd}) = \delta^{ab}_{cd} , \quad V^a (P_b) = \delta^a_b , \quad B^{ab}(Z_{cd}) = \delta^{ab}_{cd}, \nonumber \\
& \psi (Q) = \mathbf{1} , \quad \xi (\Sigma) = \mathbf{1} .
\end{align}
The aforementioned Maurer-Cartan equations read
\begin{align}
& d \omega^{ab} + \omega^{ac} \wedge \omega_{c}^{\; b}=0, \\
& D V^a - \frac{1}{2}\bar{\psi} \wedge \gamma^a \psi =0, \\
& D \psi =0, \\
& D B^{ab} + \bar{\xi}\wedge \gamma^{ab} \psi + V^a \wedge V^b=0, \label{smeqbab} \\
& D \xi - \frac{1}{2} \gamma_a \psi \wedge V^a =0, \label{smeqxi}
\end{align}
where $D=d + \omega$ denotes the Lorentz covariant derivative in four dimensions and where $\wedge$ is the wedge product between differential forms. All spinors above are Majorana spinors. The $1$-form fields of (the dual Maurer-Cartan formulation of) the super-Maxwell algebra have dimensions $[\omega^{ab}] = L^0$, $[V^a] = L$, $[\psi] = L^{1/2}$, $[B^{ab}]= L^2$, and $[\xi]=L^{3/2}$.

Let us now formulate the $\mathcal{N}=1$, $D=4$ supergravity theory in a geometric superspace approach\footnote{In this
context, the bosonic vielbein $V^a$ together with the gravitino $1$-form $\psi$ span a basis of the cotangent superspace $K=\lbrace V^a, \psi \rbrace$, where also the superspace $p$-forms (whose pull-back on space-time corresponds to $p$-index antisymmetric tensors) are defined.}, in which we can write a supersymmetric FDA involving a $2$-form potential $A^{(2)}$. Explicitly, the supersymmetric FDA defining the ground state (\textit{i.e.} the ``vacuum'') of this model is given by the
vanishing of the following set of supercurvatures:
\begin{align}
& \mathcal{R}^{ab} \equiv d \omega^{ab} + \omega^{ac}\wedge \omega_{c}^{\; b} =0, \label{fdafirst} \\
& R^a \equiv D V^a - \frac{1}{2}\bar{\psi} \wedge \gamma^a \psi =0, \\
& \rho \equiv D \psi =0, \label{dpsifda} \\
& F^{(3)} \equiv dA^{(2)} - \frac{1}{2}\bar{\psi}\wedge \gamma_a \psi \wedge V^a =0 .  \label{fdalast}
\end{align}
The $d^2$-closure of the above FDA relies in the Fierz identity (\ref{F1}) of Appendix \ref{app}.

Let us mention that the interacting theory (that is to say, out of the ground state) is obtained by introducing a non-vanishing value for the supercurvatures (defined in the left-hand side of the FDA). We will not further elaborate on the theory out of the vacuum in the present paper. We will concentrate, instead, on the cohomological structure of the theory, which is fully captured by the ground state FDA.

Now, one could wonder whether the FDA structure (\ref{fdafirst})-(\ref{fdalast}) can be traded with an ordinary Lie superalgebra written in terms of $1$-form gauge fields valued in non-trivial tensor representations of the Lorentz group (on the same lines of the study that was carried on in \cite{D'AuriaFre} and recalled and further analyzed in \cite{Hidden} in the case of $D=11$ supergravity). Observe that this cannot be done without introducing further $1$-form fields in the theory.

On the other hand, interestingly, in the present case this can be done by considering the extra fields (naturally) appearing in the Maxwell superalgebra, namely by introducing in the FDA describing the theory also the Maurer-Cartan equations (\ref{smeqbab}) and (\ref{smeqxi}).

Indeed, if we consider the following decomposition of the $2$-form $A^{(2)}$ in terms of $1$-forms (that, in this case, is also the most general one we can write provided the FDA structure above and satisfying the Bianchi identity in superspace of the $2$-form, $d^2 A^{(2)} = 0$):
\begin{equation}\label{par2}
A^{(2)}(\sigma) = \alpha \bar{\psi} \wedge \xi ,
\end{equation}
being $\alpha$ a free parameter, we have that (\ref{par2}) enjoys the ground state FDA requirement $d A^{(2)} = \frac{1}{2} \bar{\psi} \wedge \gamma_a \psi \wedge V^a$ (see (\ref{fdalast})) if
\begin{equation}\label{par2fixed}
A^{(2)}(\sigma) = - \bar{\psi} \wedge \xi ,
\end{equation}
that is to say $\alpha=-1$, where, in particular, we have used the Maurer-Cartan equation
\begin{equation}
D \xi = \frac{1}{2} \gamma_a \psi \wedge V^a . \label{dxiiiiiiiii}
\end{equation}

Observe that the $1$-form field $B^{ab}$ does not appear in the parametrization of $A^{(2)}$, where the crucial role is played just by the extra spinor $1$-form field $\xi$ appearing in the super-Maxwell algebra. Indeed, one could have obtained the same result by simply considering a (Lorentz-valued) central spinor extension (given by (\ref{dxiiiiiiiii})) of the super-Poincar\'{e} algebra in $D=4$. However, the peculiarity of our result lies in the fact that the spinor $\xi$ that allows to write the supersymmetryzation of the $D=4$ Maxwell algebra (in its dual Maurer-Cartan formulation) is also the same spinor that allows to write the parametrization (\ref{par2fixed}) in terms of $1$-forms for the $2$-form $A^{(2)}$ appearing in the FDA of the $\mathcal{N}=1$, $D=4$ supergravity theory;\footnote{As we will see in the sequel, this will also hold for the higher-dimensional case of the $D=11$ FDA describing $D=11$ supergravity.} then, in light of this fact, even if the $1$-form field $B^{ab}$ is ruled out by the parametrization of $A^{(2)}$, its contribution at the algebraic level cannot be discarded, being $B^{ab}$ a $1$-form field of (the dual Maurer-Cartan formulation of) the Maxwell superalgebra. In particular, it could be related to a possible enhancement of the (hidden gauge) symmetries underlying supergravity models, for example when considering extensions or expansions of the super-Maxwell algebra.

Let us mention, here, that we could also have added to the FDA (\ref{fdafirst})-(\ref{fdalast}) the following equation:
\begin{equation} \label{xi2giga}
D \Xi^{(2)} + \psi \wedge A^{(2)} - \frac{1}{2} \gamma_{ab}\psi \wedge V^a \wedge V^b =0,
\end{equation}
being $\Xi ^{(2)}$ a spinor $2$-form whose dimension is $[\Xi^{(2)}]=L^{5/2}$ (see also Ref. \cite{Castellani}). However, in this case, in order to write the $2$-form $\Xi^{(2)}$ in terms of $1$-forms, we would need extra $1$-form fields with respect to those appearing in the dual Maurer-Cartan formulation of the Maxwell superalgebra (for instance, $1$-form fields coming from extensions or expansions of the super-Maxwell algebra).\footnote{Or, directly, another hidden Lie superalgebra, different with respect to the super-Maxwell algebra, trivializing (\ref{xi2giga}) in terms of $1$-forms.} In the present paper, we limit ourselves to consider the supersymmetric FDA containing the $2$-form $A^{(2)}$ and leave the analysis of the FDA involving (\ref{xi2giga}) to future investigations.

From the above result, we can conclude that the super-Maxwell algebra written in (\ref{smlast}) can be interpreted as a hidden superalgebra underlying the $D=4$ supersymmetric FDA describing $\mathcal{N}=1$, $D=4$ supergravity extended to include a $2$-form $A^{(2)}$. 

Let us recall that the inclusion of a new $p$-form (a gauge potentials enjoying a gauge freedom) in the basis of the so-called $\mathcal{H}$-relative Chevalley-Eilenberg (CE) cohomology of a FDA is physically meaningful only if the whole of the FDA is gauge invariant, and this requires the non-physical degrees of freedom to be projected out from the FDA (see Ref. \cite{Hidden} for details). 

Thus, we now move to the analysis of the hidden gauge structure of the supersymmetric ground state FDA (\ref{fdafirst})-(\ref{fdalast}), on the same lines of \cite{Hidden}.

\subsection{Analysis of the hidden gauge structure in D=4}

In the following, we analyze in detail the hidden gauge structure of the FDA (\ref{fdafirst})-(\ref{fdalast}) when the $2$-form $A^{(2)}$ is parametrized in terms of $1$-forms. In particular, we investigate the conditions under which the gauge invariance of the FDA is realized once $A^{(2)}$ is expressed in terms of $1$-forms. 

In the geometrical approach adopted in this work, the fields are naturally defined in an enlarged manifold corresponding to the supergroup-manifold, where all the invariances of the FDA are diffeomorphisms generated by Lie derivatives. The physical request that the FDA should be described in term of fields living in ordinary superspace corresponds to require the Lie superalgebra to have a fiber bundle structure, whose base space is spanned by the supervielbein, the rest of the fields spanning a fiber $\mathcal{H}$. This in turn implies that the gauge fields belonging to $\mathcal{H}$ must be excluded from the construction of the so-called cochains (corresponding to gauge invariance). In geometrical terms, this corresponds to require the CE cohomology to be restricted to the $\mathcal{H}$-relative CE cohomology (see Ref. \cite{Hidden} for details).

Once the supersymmetric ground state FDA (\ref{fdafirst})-(\ref{fdalast}) is parametrized in terms of $1$-forms, the symmetry structure is based on the hidden supergroup-manifold $G$ having the structure of a principal fiber bundle $(G/\mathcal{H},\mathcal{H})$, where $G/ \mathcal{H}$ corresponds to superspace and where the fiber $\mathcal{H}$ includes, in the present case, the Lorentz transformations and the hidden super-Maxwell generators $Z_{ab}$ and $\Sigma$.

Explicitly, we can write $\mathcal{H} = H_0 + H_b + H_f$, where $\lbrace J_{ab} \rbrace \in H_0$, $ \lbrace Z_{ab} \rbrace \in H_b$, $\lbrace \Sigma \rbrace \in H_f$, and $\lbrace P_a , Q \rbrace \in \mathbb{K}$; $\mathbb{G} = \mathcal{H} + \mathbb{K}$ is the hidden Maxwell superalgebra.\footnote{With an abuse of notation, here and in the following we will use for the cotangent space of the supergroup-manifold $G$, spanned by the $1$-forms, the same symbols defined above for the tangent space of $G$, spanned by the vector fields (generators).} Observe that the subalgebra $H_b + H_f$ defines an abelian ideal of $\mathbb{G}$.

Requiring the physical condition that the CE cohomology is restricted to the $\mathcal{H}$-relative CE cohomology corresponds, now, to require the FDA to be described in terms of $1$-form fields living on $G/\mathcal{H}$; this implies that the $1$-forms in $H_b$ and $H_f$ do not appear in $dA^{(2)}$.

Now, taking into account this discussion, we consider in detail the relation between the gauge transformations of the FDA and those of the super-Maxwell bosonic and fermionic $1$-forms $B^{ab}$ and $\xi$, respectively.
The FDA (\ref{fdafirst})-(\ref{fdalast}) is invariant under the following gauge transformation:
\begin{equation}\label{gaugefda}
\delta A^{(2)}=d\Lambda^{(1)} ,
\end{equation}
which is generated by the arbitrary $1$-form $\Lambda^{(1)}$.

The gauge transformations of the bosonic 1-form $B^{ab}$ and of the spinor $1$-form $\xi$ generated by the tangent vectors in $H_b$ and in $H_f$ are
\begin{equation}\label{gauge1foms}
\left\{
\begin{array}{l}
\delta B^{ab}=d\Lambda^{ab} - \bar{\varrho} \gamma^{ab} \psi , \\
\delta \xi = D \varrho ,
\end{array}\right.
\end{equation}
where $\Lambda^{ab}$ is an arbitrary Lorentz-valued scalar function (\textit{i.e.} a $0$-form) and where we have introduced the infinitesimal spinor parameter $\varrho$. Observe that the parameter $\varrho$ appears in both the gauge transformations, while $\delta \xi$ does not involve $\Lambda^{ab}$, in agreement with the fact that the covariant
differential $D\xi$ (see equation (\ref{dxiiiiiiiii})) is parametrized only in terms of the supervielbein and not in terms of $B^{ab}$ in $H_b$. This is different from what happened in the case of the hidden superalgebra underlying the FDA of $D=11$ supergravity \cite{D'AuriaFre}, \cite{Hidden}, where the covariant differential of the spinor $1$-form dual to the extra, nilpotent, fermionic generator is parametrized also in terms of the gauge fields in $H_b$.

In the present case, the corresponding $1$-form gauge parameter of $A^{(2)}$ turns out to be given by
\begin{equation} \label{l1}
\Lambda^{(1)} = \bar{\psi} \varrho ,  
\end{equation}
where we have used the relation $\alpha=-1$ which must be fulfilled in order for (the differential of) the parametrization of $A^{(2)}$ to be equivalent to (\ref{fdalast}).

We can now show that all the diffeomorphisms in the hidden supergroup $G$, generated by Lie derivatives, are invariances of the FDA, and that, in particular, the ones in the fiber $\mathcal{H}$ directions are associated with a particular form of the gauge parameter of the FDA
gauge transformation given by (\ref{gaugefda}). Indeed, defining the tangent vectors\footnote{Since the Lorentz transformations, belonging to $H_0 \subset \mathcal{H}$, are not effective on the FDA, the $2$-form $A^{(2)}$ being Lorentz-invariant, our analysis reduces to consider the transformations induced by the tangent vectors $Z_{ab} \in H_b \subset \mathcal{H}$ and $ \Sigma \in H_f \subset \mathcal{H}$.}
\begin{align}
& \overrightarrow{z} \equiv \Lambda^{ab} Z_{ab} \in H_b , \\
& \overrightarrow{q} \equiv \bar{\varrho} \Sigma \in H_f ,
\end{align}
we find that a gauge transformation leaving invariant the FDA (\ref{fdafirst})-(\ref{fdalast}) is recovered, when $A^{(2)}$ is parametrized in terms of $1$-forms, if
\begin{equation}
\Lambda^{(1)} \equiv \Lambda^{(1)}_b + \Lambda^{(1)}_f = \imath_{\overrightarrow{z}} (A^{(2)}) + \imath_{\overrightarrow{q}} (A^{(2)}) ,
\end{equation}
where $\imath$ denotes the contraction operator and where we have denoted by $\Lambda^{(1)}_b$ the $1$-form gauge parameter corresponding to the transformation in $H_b$, while $\Lambda^{(1)}_f$ is the $1$-form gauge parameter corresponding to the transformation in $H_f$. Note that, since $\Lambda^{(1)}_b=\imath_{\overrightarrow{z}} (A^{(2)})=0$, we can write $\Lambda^{(1)} =\Lambda^{(1)}_f=  \imath_{\overrightarrow{q}} (A^{(2)})$. 

Now, introducing the Lie derivative $\ell_{\overrightarrow{z}} \equiv d \imath_{\overrightarrow{z}} + \imath_{\overrightarrow{z}} d$ (and, analogously, $\ell _{\overrightarrow{q}} \equiv d \imath_{\overrightarrow{q}} + \imath_{\overrightarrow{q}} d$), we find the corresponding gauge transformations of $A^{(2)}$ to be
\begin{align}
& \delta _{\overrightarrow{z}} A^{(2)} = 0 = d \left(\imath_{\overrightarrow{z}} (A^{(2)}) \right) = \ell_{\overrightarrow{z}} A^{(2)} , \\
& \delta _{\overrightarrow{q}} A^{(2)} = - \bar{\psi} \wedge D \varrho = d \left(\imath_{\overrightarrow{q}} (A^{(2)}) \right) = \ell_{\overrightarrow{q}} A^{(2)}.
\end{align}
The last equality in both the above relations follows since $dA^{(2)}$, as given in (\ref{fdalast}), is invariant under transformations
generated by $\overrightarrow{z}$ and $\overrightarrow{q}$ corresponding to the gauge invariance of the supervielbein. In particular, this is in agreement with the fact that the right hand side of $dA^{(2)}$ as given in (\ref{fdalast}) is in the $\mathcal{H}$-relative CE cohomology. 

Thus, after integration by parts, we can finally write:
\begin{equation}
\delta A^{(2)} = \delta _{\overrightarrow{z}} A^{(2)} + \delta _{\overrightarrow{q}} A^{(2)} = d\Lambda^{(1)}_b + d \Lambda^{(1)}_f = d \Lambda^{(1)} ,
\end{equation}
which, due to the fact that $ \delta _{\overrightarrow{z}} A^{(2)}=d\Lambda^{(1)}_b=0$, reduces to
\begin{equation}
\delta A^{(2)} =  \delta _{\overrightarrow{q}} A^{(2)} = d \Lambda^{(1)}_f = d \Lambda^{(1)} .
\end{equation}

We have thus completed the analysis of the hidden gauge structure of the $D=4$ FDA (\ref{fdafirst})-(\ref{fdalast}). We now extend our study to the $D=11$ FDA describing the Cremmer-Julia-Scherk $D=11$ supergravity theory \cite{Cremmer}.

\section{Maxwell superalgebra and hidden gauge structure of the supergravity FDA in D=11} \label{11d}

In this section, we move to the analysis of the FDA describing the $D=11$ supergravity theory of \cite{Cremmer}. In particular, we will see that, also in this case, there exists a super-Maxwell algebra which can be interpreted as a (hidden) superalgebra underlying the theory.

The $D=11$ supergravity theory, whose action was first constructed in \cite{Cremmer}, has a bosonic field content given by the metric $g_{\mu\nu}$ and a $3$-index antisymmetric tensor $A_{\mu\nu\rho}$ ($\mu,\nu,\rho,\ldots =0,1,\ldots ,D-1$); the theory is also endowed with a single Majorana gravitino $\Psi_\mu$ in the fermionic sector.\footnote{We denote by $\Psi$ the gravitino in $D=11$ space-time dimensions, in order to avoid confusion with the gravitino $\psi$ of the four-dimensional case.} By dimensional reduction, the $D=11$ theory yields $\mathcal{N}=8$, $D=4$ supergravity, which is considered as a possible unifying theory of all interactions. 

In the FDAs framework, the bosonic sector of the theory includes, besides the supervielbein $\lbrace V^a,\Psi \rbrace$ (where $a=0,1, \ldots, 10$ and where $\Psi$ is a $32$-components Majorana spinor), a $3$-form potential $A^{(3)}$ (whose pull-back on space-time is $A_{\mu \nu \rho}$), with field-strength $F^{(4)}= dA^{(3)}$ (modulo fermionic bilinears in terms of the gravitino $1$-form), together with its Hodge dual $F^{(7)}$ (whose space-time components are related to the ones of the $4$-form by $F_{\mu_1    \ldots       \mu_7}= \frac 1{84} \epsilon_{\mu_1   \ldots       \mu_7\nu_1   \ldots       \nu_4} F^{\nu_1   \ldots       \nu_4}$) associated with a $6$-form potential $B^{(6)}$ in superspace (see Ref. \cite{D'AuriaFre} for details on the FDA formulation of $D=11$ supergravity in the superspace geometric approach). 

As we have already mentioned, in \cite{D'AuriaFre} the supersymmetric FDA describing $D=11$ supergravity was introduced and then interpreted in terms of an ordinary Lie superalgebra. The superalgebra found by the authors of \cite{D'AuriaFre} can also be viewed as a spinor central extension of the so-called $M$-algebra \cite{Sezgin:1996cj}, \cite{deAzcarraga:1989mza}, \cite{Townsend:1997wg}, \cite{Hassaine:2003vq}, \cite{Hassaine:2004pp}.

In particular, the authors of \cite{D'AuriaFre} presented a general decomposition of the $3$-form $A^{(3)}$ in terms of $1$-forms, by requiring the Bianchi identity in superspace of the $3$-form, $d^2A^{(3)} = 0$, to be satisfied also when $A^{(3)}$ is written in terms of $1$-forms. The result of \cite{D'AuriaFre} (the authors got a dichotomic solution, consisting in two different supergroups, whose $1$-form potentials can be alternatively used to parametrize the $3$-form) have been further analyzed and generalized in \cite{Hidden}, \cite{Bandos:2004xw}, \cite{Bandos:2004ym}, where some misprints of \cite{D'AuriaFre} have been corrected and where, in particular in \cite{Bandos:2004xw} and \cite{Bandos:2004ym}, it was pointed out that a restriction imposed in \cite{D'AuriaFre} on one coefficient in the parametrization of $A^{(3)}$ can be relaxed, thus giving a one-parameter family of solutions.

In the following, we will see that there also exists another hidden Lie superalgebra underlying the FDA describing $D=11$ supergravity (we will not consider the complete $D=11$ FDA involving a $6$-form potential $B^{(6)}$ in the present work, limiting ourselves to the FDA containing just a $3$-form $A^{(3)}$), namely a minimal Maxwell superalgebra in eleven dimensions. In particular, we will see that the general parametrization of $A^{(3)}$ in terms of the hidden super-Maxwell $1$-forms, together with $V^a$ and $\Psi$, can be recast into the form of that written in Refs. \cite{Bandos:2004xw}, \cite{Bandos:2004ym}.

The supersymmetric FDA defining the ground state of the $D=11$ theory\footnote{We do not consider the $D=11$ theory out of the vacuum in the present paper. Some progress in this topic has been obtained in Ref. \cite{Bandos:2004ym}.} is given by the vanishing of the following supercurvatures:
\begin{align}
& \mathcal{R}^{ab} \equiv d\omega^{ab} -  \omega^{ac}\wedge \omega_c^{\; b}=0 ,\label{FDA11omega} \\
& R^a \equiv D V^a - \frac{\ii}{2}\bar{\Psi}\wedge \Gamma^a \Psi =0 ,\label{FDA11v} \\
& \rho \equiv D \Psi = 0 ,\label{FDA11psi}\\
& F^{(4)} \equiv dA^{(3)} - \frac{1}{2}\bar{\Psi}\wedge \Gamma_{ab}\Psi \wedge V^a \wedge V^b =0 , \label{FDA11a3} 
\end{align}
where $D$ ($D=d-\omega$, according with the convention of \cite{D'AuriaFre}, \cite{Hidden}) denotes the Lorentz covariant derivative in eleven dimensions and where $\Gamma_a$ and $\Gamma_{ab}$ are gamma matrices in $D=11$. Again, the vielbein $V^a$ and the gravitino $\Psi$ span a basis of the cotangent superspace $K \equiv \lbrace V^a,\Psi \rbrace$ where also the superspace $3$-form $A^{(3)}$ is defined. The $d^2$-closure of the FDA (\ref{FDA11omega})-(\ref{FDA11a3}) is a consequence of $3$-gravitinos Fierz identity $\Gamma_{ab} \Psi \wedge \bar{\Psi} \wedge \Gamma^a \Psi =0$ in $D=11$.

Let us now consider the following minimal Maxwell superalgebra (written in its dual Maurer-Cartan formulation) in eleven dimensions:
\begin{align}
& d \omega^{ab} - \omega^{ac} \wedge \omega_{c}^{\; b}=0, \label{domprima} \\
& D V^a = \frac{\ii}{2}\bar{\Psi} \wedge \Gamma^a \Psi , \\
& D \Psi =0, \\
& D \tilde{B}^{ab} = \frac{1}{2} \bar{\Psi} \wedge \Gamma^{ab}\Psi , \label{maxtildebab} \\
& D B^{ab} = - \bar{\chi}\wedge \Gamma^{ab} \Psi - V^a \wedge V^b - \frac{1}{5} \tilde{B}^{ac} \wedge \tilde{B}_c^{\; b} , \label{maxbab} \\
& D \chi = \frac{\ii}{2} \gamma_a \Psi \wedge V^a - \frac{1}{20}\gamma_{ab} \Psi \wedge \tilde{B}^{ab} , \label{maxxi}
\end{align}
where $\chi$ is a spinor $1$-form (with dimension $[\chi]=L^{3/2}$) dual to a nilpotent fermionic generator (we have used the symbol $\chi$ in order to avoid confusion with the spinor $1$-form $\xi$ of the four-dimensional case discussed in Section \ref{Maxalg}). The $d^2$-closure of this superalgebra is a consequence of $3$-gravitinos Fierz identities in $D=11$ (see Appendix \ref{app}).

Observe that the generator dual to the $1$-form field $\tilde{B}^{ab}$, let us call it $\tilde{Z}_{ab}$, is a non-abelian one. In the absence of the super-Maxwell fields, this bosonic generator would become an almost-central bosonic generator; in eleven dimensions, it was understood as a $2$-brane charge, source of a $3$-form gauge potential (see, for example, Ref. \cite{Hidden} for details). The (dual Maurer-Cartan formulation of the) superalgebra (\ref{domprima})-(\ref{maxxi}) is a $D=11$ extension including an extra bosonic $1$-form field $\tilde{B}^{ab}$ (whose dimension is $[\tilde{B}^{ab}]=L$) of the $D=4$ super-Maxwell algebra we have considered in Section \ref{Maxalg}. Note that the superalgebra (\ref{domprima})-(\ref{maxxi}) have the same form of the minimal super-Maxwell algebra in $D=4$ discussed in \cite{CR2} (which was referred to as $s\mathcal{M}_4$ in that paper).\footnote{However, the $1$-form fields of \cite{CR2} have different dimensions with respect to those appearing in (\ref{domprima})-(\ref{maxxi}).}

Now, the most general ansatz for the $3$-form $A^{(3)}$, written in terms of the $1$-forms $\sigma^A= \lbrace V^a, \tilde{B}^{ab}, B^{ab}, \Psi, \chi \rbrace$, satisfying the Bianchi identity $d^2 A^{(3)}=0$ reads as follows:
\begin{align}
A^{(3)} (\sigma) &=  T_0 \tilde{B}^{ab} \wedge V_a \wedge V_b + T_1 \tilde{B}^{ab} \wedge \tilde{B}_{bc} \wedge \tilde{B}^{c}_{\; a} + \nonumber \\
& \quad + \ii S_1 \bar{\Psi} \wedge \Gamma_a \chi \wedge V^a + S_2 \bar{\Psi} \wedge \gamma_{ab} \chi \wedge \tilde{B}^{ab} + \nonumber \\
& \quad + M_1 \bar{\Psi} \wedge \Gamma_{ab} \Psi \wedge B^{ab}. \label{a3par}
\end{align}
Then, the requirement that expression $A^{(3)}(\sigma)$ in (\ref{a3par}) satisfies the FDA equation (\ref{FDA11a3}) leads to the following system of equations involving the coefficients $T_0$, $T_1$, $S_1$, $S_2$, and $M_1$:
\begin{equation} \label{cond11}
\left\{
\begin{array}{l}
T_0- S_1 -2M_1 -1 =0 , \cr
T_0+ \frac{1}{10}S_1 -S_2 =0 , \cr
\frac{3}{2} T_1 + \frac{1}{5} S_2 + \frac{1}{5} M_1 =0 , \cr
- \frac{1}{2} S_1 - 5 S_2 + 10 M_1=0 .
\end{array}
\right.
\end{equation}
The solution to the system (\ref{cond11}) depends on one free parameter (we choose $M_1$), and it is given by:
\begin{align}
& T_0 = \frac{1}{6} + 2M_1, \quad T_1 = - \frac{1}{90}- \frac{2}{5}M_1 , \nonumber \\
& S_1=- \frac{5}{6}, \quad S_2 = \frac{1}{12} + 2 M_1. \label{solution}
\end{align}

Some remarks are in order. First of all, one can easily prove that in the absence of the super-Maxwell extra spinor $\chi$ the expression (\ref{a3par}) could not reproduce the FDA equation (\ref{FDA11a3}) on ordinary superspace anymore. On the other hand, using equation (\ref{maxtildebab}), the last term in (\ref{a3par}) can be rewritten as
\begin{equation}
M_1 \bar{\Psi} \wedge \Gamma_{ab} \Psi \wedge B^{ab}=2 M_1 D\tilde{B}_{ab}\wedge B^{ab} .
\end{equation}
Then, we have
\begin{equation}
2 M_1 D\tilde{B}_{ab}\wedge B^{ab} = 2M_1 d(\tilde{B}_{ab} \wedge B^{ab})+2M_1 \tilde{B}_{ab} \wedge D B^{ab}
\end{equation}
and, extracting the total derivative (which is allowed since the FDA is invariant under the $3$-form gauge transformation $\delta A^{(3)}=d\Lambda^{(2)}$) and using equation (\ref{maxbab}), we obtain the following expression for $A^{(3)}$ in terms of $1$-forms:
\begin{align}
A^{(3)} &=  \left(T_0 - 2 M_1 \right) \tilde{B}^{ab} \wedge V_a \wedge V_b + \nonumber \\
& \quad + \left( T_1+ \frac{2}{5} M_1 \right) \tilde{B}^{ab} \wedge \tilde{B}_{bc} \wedge \tilde{B}^{c}_{\; a} + \nonumber \\
& \quad + \ii S_1 \bar{\Psi} \wedge \Gamma_a \chi \wedge V^a + \left( S_2 - 2 M_1 \right) \bar{\Psi} \wedge \gamma_{ab} \chi \wedge \tilde{B}^{ab}. \label{a3parNEEEEEW}
\end{align}
This final expression contains only the terms appearing in the composite $3$-form written in Refs. \cite{Bandos:2004xw}, \cite{Bandos:2004ym}. In particular, it does not contain the bosonic $1$-form field $B^{ab}$. Accordingly, redefining $T_0 - 2 M_1  \equiv \hat{T}_0$, $ T_1+ \frac{2}{5} M_1 = \hat{T}_1$, $S_1 \equiv \hat{S}_1$, and $S_2 - 2 M_1 \equiv \hat{S}_2$ in (\ref{a3parNEEEEEW}) (that is equivalent to set $M_1=0$ in (\ref{a3par})) and imposing the requirement that the expression for $A^{(3)}$ in (\ref{a3parNEEEEEW}) satisfies the FDA equation (\ref{FDA11a3}), one ends up with 
\begin{equation}
\hat{T}_0 = \frac{1}{6} , \quad \hat{T}_1 = - \frac{1}{90}, \quad \hat{S}_1=- \frac{5}{6}, \quad \hat{S}_2 = \frac{1}{12} ,
\end{equation}
corresponding to the solution found in \cite{Bandos:2004xw}, \cite{Bandos:2004ym} with a particular choice for the normalization of the extra spinor $1$-form (see also \cite{Hidden} and, in particular, the expression for $A^{(3)}_{(0)}$ in \cite{Malg} where, however, the extra spinor $1$-form named $\xi$ was normalized in a different way).\footnote{In the case under analysis, we are not considering the presence of the extra bosonic $1$-form field $B^{a_1 \ldots a_5}$ (dual to a bosonic generator $Z_{a_1 \ldots a_5}$), which would appear when considering the complete FDA including also a $6$-form potential $B^{(6)}$.} 

Thus, we can conclude that the super-Maxwell algebra (\ref{domprima})-(\ref{maxxi}) can be interpreted as a (hidden) superalgebra underlying the supersymmetric FDA (\ref{FDA11omega})-(\ref{FDA11a3}) describing $D=11$ supergravity. This superalgebra is larger than the one discovered in \cite{D'AuriaFre} (excluding the $1$-form $B^{a_1 \ldots a_5}$), since in contains one more extra bosonic $1$-form field $B^{ab}$. 
On the other hand, the contribution coming from $B^{ab}$ in the parametrization of $A^{(3)}$ can be reabsorbed by a gauge transformation of the $3$-form. Again, in analogy with the result we have obtained in Section \ref{Maxalg} in $D=4$ space-time dimensions, the peculiarity of the above result in $D=11$ lies in the fact that the spinor $\chi$ that allows to write the supersymmetryzation of the $D=11$ Maxwell algebra is also the same spinor that allows to write the parametrization of the $3$-form $A^{(3)}$ in terms of $1$-forms in such a way to fulfill the FDA requirement (\ref{FDA11a3}). 

We now move to the analysis of the hidden gauge structure of the supersymmetric FDA (\ref{FDA11omega})-(\ref{FDA11a3}).

\subsection{Analysis of the hidden gauge structure in D=11}

Recalling the discussion presented in Section \ref{Maxalg}, once the supersymmetric FDA (\ref{FDA11omega})-(\ref{FDA11a3}) is parametrized in terms of $1$-forms, the symmetry structure is based on the hidden supergroup-manifold $G$ having the structure of a principal fiber bundle $(G/\mathcal{H},\mathcal{H})$: $G/ \mathcal{H}$ corresponds to superspace, while the fiber $\mathcal{H}$ in the present $D=11$ case includes, besides the Lorentz transformations, also the hidden super-Maxwell generators in $D=11$ (we call them $\tilde{Z}_{ab}$, $Z_{ab}$, and $\Sigma$, and they are dual to the $1$-form fields $\tilde{B}^{ab}$, $B^{ab}$, and $\chi$, respectively).

We can then write $\mathcal{H} = H_0 + H_b + H_f$, so that $\lbrace J_{ab} \rbrace \in H_0$, $ \lbrace \tilde{Z}_{ab}, Z_{ab} \rbrace \in H_b$, $\lbrace \Sigma \rbrace \in H_f$, and $\lbrace P_a , Q \rbrace \in \mathbb{K}$, where $\mathbb{G} = \mathcal{H} + \mathbb{K}$ is the hidden Maxwell superalgebra.

We now analyze the relation between the FDA gauge transformations and those of its hidden Maxwell supergroup. As we have already mentioned, the FDA (\ref{FDA11omega})-(\ref{FDA11a3}) is invariant under the gauge transformation
\begin{equation}\label{gaugefdaNEW}
\delta A^{(3)}=d\Lambda^{(2)} ,
\end{equation}
which is generated by the arbitrary $2$-form $\Lambda^{(2)}$.

The gauge transformations of the bosonic 1-forms $\tilde{B}^{ab}$, $B^{ab}$ and of the spinor $1$-form $\chi$ generated by the tangent vectors in $H_b$ and in $H_f$ are respectively given by:
\begin{equation}\label{gauge1fomsNEW}
\left\{
\begin{array}{l}
\delta \tilde{B}^{ab} = d \tilde{\Lambda}^{ab} , \\
\delta B^{ab}=d\Lambda^{ab} - \bar{\varrho} \gamma^{ab} \Psi - \frac{2}{5} \tilde{\Lambda}^{ac} \tilde{B}_c^{\;b}, \\
\delta \chi = D \varrho + \frac{1}{20} \Gamma_{ab} \Psi \tilde{\Lambda}^{ab},
\end{array}\right.
\end{equation}
where $\tilde{\Lambda}^{ab}$ and $\Lambda^{ab}$ are arbitrary Lorentz-valued scalar functions and where we have introduced the infinitesimal spinor parameter $\varrho$.
The corresponding $2$-form gauge parameter of $A^{(3)}$ turns out to be
\begin{align}
\Lambda^{(2)} &= T_0 \tilde{\Lambda}^{ab} V_a \wedge V_b + 3T_1 \tilde{\Lambda}^{ab} \tilde{B}_{bc} \wedge \tilde{B}^c_{\;a} + \nonumber \\
& \quad - \ii S_1 \bar{\Psi} \wedge \Gamma_a \varrho V^a - S_2 \bar{\Psi} \wedge \Gamma_{ab} \varrho \tilde{B}^{ab} + \nonumber \\
& \quad +S_2 \bar{\Psi} \wedge \Gamma_{ab} \chi \tilde{\Lambda}^{ab} +M_1 \bar{\Psi} \wedge \Gamma_{ab} \Psi \Lambda^{ab} .  \label{l2NEW}
\end{align}

We can now show that all the diffeomorphisms in the hidden Maxwell supergroup, generated by Lie derivatives, are invariances of the FDA, the ones in the fiber $\mathcal{H}$ directions being associated with a particular form of the gauge parameter of the FDA
gauge transformation (\ref{gaugefdaNEW}). Indeed, defining the following tangent vectors:\footnote{Again, since the Lorentz transformations, belonging to $H_0 \subset \mathcal{H}$, are not effective on the FDA, the $3$-form $A^{(3)}$ being Lorentz-invariant, our analysis reduces to consider the transformations induced by the tangent vectors in $H_b $ and in $ H_f$.}
\begin{align}
& \overrightarrow{z} \equiv \tilde{\Lambda}^{ab} \tilde{Z}_{ab} + \Lambda^{ab} Z_{ab} \in H_b , \\
& \overrightarrow{q} \equiv \bar{\varrho} \Sigma \in H_f ,
\end{align}
we find that a gauge transformation leaving invariant the FDA (\ref{FDA11omega})-(\ref{FDA11a3}) is recovered, $A^{(3)}$ being parametrized in terms of $1$-forms, if
\begin{equation}
\Lambda^{(2)} \equiv \Lambda^{(2)}_b + \Lambda^{(2)}_f = \imath_{\overrightarrow{z}} (A^{(3)}) + \imath_{\overrightarrow{q}} (A^{(3)}) ,
\end{equation}
where $\imath$ denotes the contraction operator and where we have denoted by $\Lambda^{(2)}_b$ the $2$-form gauge parameter corresponding to the transformations in $H_b$, while $\Lambda^{(2)}_f$ is the $2$-form gauge parameter corresponding to the transformation in $H_f$. The result written above is true as a consequence of the relations (\ref{cond11}) obeyed by the coefficients of the parametrization (\ref{a3par}) of $A^{(3)}$ in terms of $1$-forms.

Then, introducing the Lie derivative $\ell _{\overrightarrow{z}} \equiv d \imath_{\overrightarrow{z}} + \imath_{\overrightarrow{z}} d$ (and, analogously, $\ell _{\overrightarrow{q}} \equiv d \imath_{\overrightarrow{q}} + \imath_{\overrightarrow{q}} d$), we can write
\begin{align}
\delta A^{(3)} & = \delta_{\overrightarrow{z}} A^{(3)} + \delta_{\overrightarrow{q}} A^{(3)} = \nonumber \\
& = T_0 \, d \tilde{\Lambda}^{ab} \wedge V_a \wedge V_b + 3 T_1 d \tilde{\Lambda}^{ab} \wedge \tilde{B}_{bc} \wedge \tilde{B}^c_{\; a} + \nonumber \\
& \quad + \ii S_1 \bar{\Psi} \wedge \Gamma_a \left( D \varrho + \frac{1}{20} \Gamma_{bc} \Psi \tilde{\Lambda}^{bc} \right) \wedge V^a +  \nonumber \\
& \quad  + S_2 \bar{\Psi} \wedge \gamma_{ab} \left( D \varrho + \frac{1}{20} \Gamma_{cd} \Psi \tilde{\Lambda}^{cd} \right) \wedge \tilde{B}^{ab} + \nonumber \\
& \quad + S_2 \bar{\Psi} \wedge \Gamma_{ab}\chi \wedge d \tilde{\Lambda}^{ab} + \nonumber \\
& \quad + M_1 \bar{\Psi} \wedge \Gamma_{ab} \Psi \wedge \left(d \Lambda^{ab} - \bar{\varrho} \Gamma^{ab}\Psi - \frac{2}{5}\tilde{\Lambda}^{ac} \tilde{B}_c^{\; b} \right) = \nonumber \\
& = d \left(\imath_{\overrightarrow{z}} (A^{(3)}) \right) + d \left(\imath_{\overrightarrow{q}} (A^{(3)}) \right) = \nonumber \\
& = \ell_{\overrightarrow{z}} A^{(3)} + \ell_{\overrightarrow{q}} A^{(3)} ,
\end{align}
where the last equality follows since $dA^{(3)}$, as given in (\ref{FDA11a3}), is invariant under transformations
generated by $\overrightarrow{z}$ and $\overrightarrow{q}$, corresponding to the gauge invariance of the supervielbein (the right hand side of $dA^{(3)}$ is in the $\mathcal{H}$-relative CE cohomology).

We can finally see that, after integration by parts, making use of $3$-gravitinos Fierz identities in $D=11$ (see Appendix \ref{app}) and of the relations (\ref{cond11}), the above result exactly reproduces the gauge transformation (\ref{gaugefdaNEW}) leaving invariant the supersymmetric FDA (\ref{FDA11omega})-(\ref{FDA11a3}). Precisely, we have
 \begin{equation}
\delta A^{(3)} = \delta_{\overrightarrow{z}} A^{(3)} + \delta_{\overrightarrow{q}} A^{(3)} = d \Lambda^{(2)} ,
\end{equation} 
where $\Lambda^{(2)}$ is defined in equation (\ref{l2NEW}). This result is hardly surprising, since if one had reabsorbed (as shown above) the term containing $B^{ab}$ in the parametrization (\ref{a3par}) of $A^{(3)}$, the analysis of the FDA gauge invariance would have been traced back to the one done in \cite{Hidden}.
 
We have thus completed the analysis of the hidden gauge structure of the $D=11$ supersymmetric FDA (\ref{FDA11omega})-(\ref{FDA11a3}).

\section{Conclusions} \label{Concl}

In this paper, driven by the fact that any supersymmetric FDA can always be traded for a hidden Lie superalgebra containing extra, nilpotent, fermionic generators \cite{Hidden}, we have first of all shown that the $D=4$ super-Maxwell algebra of Ref. \cite{deAzcarraga:2014jpa} (given in (\ref{smlast})) can be interpreted as a hidden superalgebra underlying the ground state FDA (\ref{fdafirst})-(\ref{fdalast}) of $D=4$ supergravity containing a $2$-form potential $A^{(2)}$.

Subsequently, we have considered the FDA (introduced in \cite{D'AuriaFre}) describing the $D=11$ supergravity theory of \cite{Cremmer}, which contains a $3$-form potential $A^{(3)}$, and we have shown that there exists a $D=11$ super-Maxwell algebra underlying the theory. In this work, we have limited ourselves to consider the $D=11$ FDA containing just the $3$-form $A^{(3)}$, leaving the study of the complete FDA involving also a $6$-form potential $B^{(6)}$ (and, correspondingly, the presence of an extra bosonic $1$-form field $B^{a_1 \ldots a_5}$, see Refs. \cite{D'AuriaFre}, \cite{Hidden}, in the underlying Lie superalgebra) to future investigations.\footnote{In that case, the extra bosonic $1$-form field $B^{ab}$ appearing in the $D=11$ super-Maxwell algebra could play a more prominent role in participating to the parametrization of $B^{(6)}$ in terms of $1$-forms.}

In the analyses we have performed, the presence of the extra spinors $\xi$ and $\chi$ naturally appearing in the supersymmetric extension of the Maxwell algebras in the $D=4$ and $D=11$ cases, respectively, is crucial in order to reproduce the $D=4$ and $D=11$ FDAs on ordinary superspace, whose basis is given by the supervielbein. Indeed, referring, for instance, to the $D=4$ case, the spinor $1$-form field $\xi$ allows to write the parametrization $A^{(2)}=-\bar{\psi} \wedge \xi$ satisfying (\ref{fdalast}); this would not be possible without adding extra fields to the $D=4$ supergravity theory, and it is particularly intriguing that it is really a fundamental spinor to the construction of the Maxwell superalgebra to make possible a parametrization in terms of $1$-forms of the $2$-form $A^{(2)}$ appearing in the FDA of the $\mathcal{N}=1$, $D=4$ supergravity theory. The same consideration holds true also in the $D=11$ case, where the extra spinor $\chi$ naturally appearing in the $D=11$ super-Maxwell algebra allows to trivialize the FDA containing the $3$-form potential $A^{(3)}$ when the latter is written in terms of $1$-forms. In this case, we have shown that, exploiting the gauge invariance of the $3$-form, the parametrization (\ref{a3par}) of $A^{(3)}$ in terms of $1$-forms can be recast into the form given in \cite{Bandos:2004xw}, \cite{Bandos:2004ym} (see also \cite{Hidden} and $A^{(3)}_{(0)}$ of \cite{Malg}). Our result could shed some light on the symmetries hidden in $D=11$ supergravity and related models (see, for instance, Refs. \cite{Bergshoeff:1995hm} and \cite{Sezgin:1996cj}). 

Concerning the $D=4$ FDA, in this work we have just considered the FDA including the $2$-form potential $A^{(2)}$. We leave the analysis of the (complete) FDA involving also a spinor $2$-form $\Xi ^{(2)}$ (see \cite{Castellani}) satisfying (\ref{xi2giga}) to future works. This would require extra $1$-form fields with respect to those appearing in the dual Maurer-Cartan formulation of the Maxwell superalgebra or, directly, a different Lie superalgebra underlying the theory.

The extra super-Maxwell fields could be important additions towards the construction of possible off-shell models underlying supergravity theories (mainly in higher-dimensional cases, such as the eleven-dimensional one). 

Furthermore, let us mention that our framework is naturally related to the formulation of Double Field Theory and Exceptional Field Theory (see also Refs. \cite{Hidden}, \cite{Malg}). Indeed, the presence of extra bosonic $1$-forms in the dual formulation of Lie superalgebras appears to be quite analogous to the presence of extra coordinate directions in the formulation of Double Field Theory and Exceptional Field Theory; in particular, referring to Exceptional Field Theory, the section constraints required in that theory to project the field equations on ordinary
superspace should be dynamically implemented through the presence of the cohomological extra spinors.

It would be interesting to extend our discussion and interpretation of the (hidden) Maxwell-superalgebras to higher-dimensional and $\mathcal{N}>1$ theories worked out in a geometric framework (also matter-coupled ones), investigating, in particular, possible supersymmetric extensions of the discussion presented in \cite{Gomis:2017cmt}. 

Finally, one could also analyze gauged FDAs in this geometric framework; in this context, we conjecture that the so-called $AdS$-Maxwell superalgebra \cite{Durka:2011gm} could play an important role within our approach. Some work is in progress on this topic.

\section*{Acknowledgements}

The author is grateful to Laura Andrianopoli and Riccardo D'Auria for the support and the stimulating discussions during the early stages of the preparation of this work. The author also wish to acknowledge illuminating discussions with Igor A. Bandos.

\appendix

\section{Notation, conventions, and useful formulas} \label{app}

In the following, we collect the conventions and some useful formulas that we have used in this work, both in $D=4$ and in $D=11$ space-time dimensions.

\subsection{Conventions and useful formulas in D=4}

The Dirac gamma matrices in $D=4$ are defined through the relation
\begin{equation}
\lbrace{ \gamma_a , \gamma_b }\rbrace = - 2 \eta_{ab},
\end{equation}
where $\eta_{ab} \equiv (-1,1,1,1)$ is the Minkowski metric.
These gamma matrices satisfy the Clifford algebra:
\begin{align}
& [\gamma_a , \gamma_b] = 2 \gamma_{ab} , \\
& \gamma_5 = - \gamma_0 \gamma_1 \gamma_2 \gamma_3 \gamma_4, \quad \quad \gamma^2_5 = -1 , \\
& \lbrace{ \gamma_5 , \gamma_a }\rbrace = [\gamma_5 , \gamma_{ab}] =0, \\
& \gamma_{ab} \gamma_5 = -\frac{1}{2}\epsilon_{abcd} \gamma^{cd}, \\
& \gamma_a \gamma_b = \gamma_{ab}-\eta_{ab}, \\
& \gamma^{ab} \gamma_{cd} = \epsilon^{ab}_{\;\;\;cd} \gamma_5 - 4 \delta^{[a}_{\;\;[c} \gamma^{b]}_{\;\;d]}- 2 \delta^{ab}_{cd}, \\
& \gamma^{ab} \gamma^c = 2 \gamma^{[a}\delta^{b]}_c - \epsilon^{abcd} \gamma_5 \gamma_d , \\
& \gamma^c \gamma^{ab} = - 2 \gamma^{[a}\delta^{b]}_c - \epsilon^{abcd}\gamma_5 \gamma_d , 
\end{align}
and
\begin{align}
& \gamma_m \gamma^{ab} \gamma^m = 0, \\
& \gamma_{ab} \gamma_m \gamma^{ab}=0, \\
& \gamma_{ab}\gamma_{cd}\gamma^{ab}= 4 \gamma_{cd}, \\
& \gamma_m \gamma^a \gamma^m = -2 \gamma^a .
\end{align}

We are working with Majorana spinors, satisying $\bar{\psi} = \psi^T C$, where $C$ is the charge conjugation matrix. Furthermore, the gamma matrices satisfy
\begin{equation}
(C \gamma_a)^T = C \gamma_a , \quad (C \gamma_{ab})^T = C \gamma_{ab},
\end{equation}
while
\begin{equation}
C^T = - C, \quad (C \gamma_5)^T = - C \gamma_5 , \quad (C \gamma_5 \gamma_a)^T = - C \gamma_5 \gamma_a ,
\end{equation}
meaning that $C \gamma_a$ and $C \gamma_{ab}$ are symmetric, while $C$, $C \gamma_5$, and $C \gamma_5 \gamma_a$ are antisymmetric gamma matrices. This leads to the following identities for the $p$-form $\psi$ and $q$-form $\xi$:
\begin{align}
& \bar{\psi} \wedge \xi = (-1)^{pq}\bar{\xi} \wedge \psi , \\
& \bar{\psi} \wedge S \xi = - (-1)^{pq} \bar{\xi} \wedge S \psi , \\
& \bar{\psi} \wedge A \xi = (-1)^{pq}\bar{\xi} \wedge A \psi ,
\end{align}
being $S$ a symmetric matrix and $A$ an antisymmetric one. 
We can then write some useful Fierz identities for $\mathcal{N}=1$, $D=4$ (for the $1$-form spinor $\psi$):
\begin{align}
& \psi \wedge \bar{\psi} = \frac{1}{2} \gamma_a \bar{\psi} \wedge \gamma^a \psi - \frac{1}{8} \gamma_{ab} \bar{\psi} \wedge \gamma^{ab} \psi , \\
& \gamma_a \psi \wedge \bar{\psi} \wedge \gamma^a \psi = 0, \label{F1} \\
& \gamma_{ab} \psi \wedge \bar{\psi} \wedge \gamma^{ab} \psi = 0, \\
& \gamma_{ab} \psi \wedge \bar{\psi} \wedge \gamma^a \psi = \psi \wedge \bar{\psi} \wedge \gamma_b \psi .
\end{align}

\subsection{Fierz identities and irreducible representations in D=11}

The gravitino $\Psi_\alpha $ (with $\alpha =1,\ldots , 32$) in $D=11$ space-time dimensions is a spinor $1$-form that belongs to the spinor representation of $ SO(1,10) \simeq Spin(32)$.

The Fierz identities amount to decompose the representation $(\alpha , \beta , \gamma)$\footnote{That is the symmetric product $(\alpha , \beta , \gamma)\equiv\Psi_{(\alpha} \wedge \Psi_\beta \wedge \Psi_{\gamma )}$, of dimension $\mathbf{5984}$, which belongs to the three-times symmetric reducible representation of $Spin(32)$.} into irreducible representations of $Spin(32)$. 

We get: $\mathbf{5984} \to \mathbf{32}+\mathbf{320}+\mathbf{1408}+\mathbf{4224}$.
Denoting the corresponding irreducible spinor representations of the Lorentz group $SO(1,10)$ as
\begin{align}
& \Xi^{(32)} \in \mathbf{32} \,,\quad \Xi^{(320)}_a \in \mathbf{320}\,, \nonumber \\
& \Xi^{(1408)}_{a_1 a_2}\in \mathbf{1408}\,,\quad \Xi^{(4224)}_{a_1    \ldots    a_5}\in \mathbf{4224} \, ,
\end{align}
where the indexes $a_1 \ldots a_n$ are antisymmetrized and where $\Gamma^a \Xi_{ab_1 \ldots b_n}=0$, one can now compute the coefficients of the explicit decomposition into the irreducible basis, obtaining (see also Refs. \cite{D'AuriaFre}, \cite{Hidden}):
\begin{align}
\Psi \wedge \bar{\Psi} \wedge \Gamma_a \Psi & = \Xi^{(320)} _a+ \frac{1}{11}\Gamma_a \Xi^{(32)},  \\
\Psi \wedge \bar{\Psi} \Gamma_{a_1 a_2}\Psi & = \Xi^{(1408)}_{a_1a_2}-\frac{2}{9}\Gamma_{[a_2} \Xi^{(320)}_{a_2]} + \nonumber \\
& \quad +\frac{1}{11}\Gamma_{a_1 a_2}\Xi^{(32)},  \\
\Psi \wedge \bar{\Psi}\wedge \Gamma_{a_1 \ldots a_5}\Psi & = \Xi^{(4224)}_{a_1    \ldots    a_5}+2 \Gamma_{[a_1 a_2 a_3}\Xi^{(1408)}_{a_4a_5]}+ \nonumber \\
& \quad + \frac{5}{9}\Gamma_{[a_1    \ldots    a_4}\Xi^{(320)}_{a_5]}- \frac{1}{77}\Gamma_{a_1    \ldots    a_5}\Xi^{(32)} .
\end{align}


\begin{thebibliography}{99}


\bibitem{Sullivan} D.~Sullivan,
 ``Infinitesimal computations in topology''. Publications Math\'{e}matiques de l'IHES, 47 (1977), p. 269-331.

\bibitem{D'AuriaFre} R.~D'Auria and P.~Fr\'{e}, 
  ``Geometric Supergravity in d = 11 and Its Hidden Supergroup,''
  Nucl.\ Phys.\ B {\bf 201} (1982) 101,
   Erratum: [Nucl.\ Phys.\ B {\bf 206} (1982) 496],
  doi:10.1016/0550-3213(82)90376-5, 10.1016/0550-3213(82)90281-4 

\bibitem{Hidden} L.~Andrianopoli, R.~D'Auria and L.~Ravera,
  ``Hidden Gauge Structure of Supersymmetric Free Differential Algebras,'' 
  JHEP {\bf 1608} (2016) 095, 
  doi:10.1007/JHEP08(2016)095
  [arXiv:1606.07328 [hep-th]].    
  
\bibitem{Cremmer} E.~Cremmer, B.~Julia and J.~Scherk,
  ``Supergravity Theory in Eleven-Dimensions,''
  Phys.\ Lett.\  {\bf 76B} (1978) 409,
  doi:10.1016/0370-2693(78)90894-8  
  
\bibitem{Hull:1994ys}
  C.~M.~Hull and P.~K.~Townsend,
  ``Unity of superstring dualities,''
  Nucl.\ Phys.\ B {\bf 438} (1995) 109,
  doi:10.1016/0550-3213(94)00559-W 
  [arXiv:9410167 [hep-th]].

\bibitem{Townsend:1995gp}
  P.~K.~Townsend,
  ``P-brane democracy,'' 
  In *Duff, M.J. (ed.): The world in eleven dimensions* 375-389,
  [arXiv:9507048 [hep-th]]. 
  
\bibitem{vanHolten:1982mx}
  J.~W.~van Holten and A.~Van Proeyen,
  ``N=1 Supersymmetry Algebras in D=2, D=3, D=4 MOD-8,''
  J.\ Phys.\ A {\bf 15} (1982) 3763,
  doi:10.1088/0305-4470/15/12/028  
  
\bibitem{Bandos:2004xw}
   I.~A.~Bandos, J.~A.~de Azcarraga, J.~M.~Izquierdo, M.~Picon and O.~Varela, 
   ``On the underlying gauge group structure of D=11 supergravity,'' 
   Phys.\ Lett.\ B {\bf 596} (2004) 145, 
   doi:10.1016/j.physletb.2004.06.079 
   [arXiv:0406020 [hep-th]].

\bibitem{Bandos:2004ym}
   I.~A.~Bandos, J.~A.~de Azcarraga, M.~Picon and O.~Varela,
   ``On the formulation of D = 11 supergravity and the composite  
nature of its three-form gauge field,''
   Annals Phys.\  {\bf 317} (2005) 238
   doi:10.1016/j.aop.2004.11.016,
   [arXiv:0409100 [hep-th]].     
  
\bibitem{Malg} 
  L.~Andrianopoli, R.~D'Auria and L.~Ravera, 
  ``More on the Hidden Symmetries of 11D Supergravity,'' 
  Phys.\ Lett.\ B {\bf 772} (2017) 578, 
  doi:10.1016/j.physletb.2017.07.016
    [arXiv:1705.06251 [hep-th]].  
  
\bibitem{SM4} D.~M.~Pe\~{n}afiel and L.~Ravera,
  ``On the Hidden Maxwell Superalgebra underlying D=4 Supergravity,'' 
  Fortsch.\ Phys.\  {\bf 65} (2017) no.9,  1700005 
  doi:10.1002/prop.201700005
  [arXiv:1701.04234 [hep-th]].  
  
\bibitem{Gunaydin:2005df}
  M.~Gunaydin, S.~McReynolds and M.~Zagermann,
  ``Unified N=2 Maxwell-Einstein and Yang-Mills-Einstein supergravity theories in four dimensions,''
  JHEP {\bf 0509} (2005) 026,
  doi:10.1088/1126-6708/2005/09/026
  [arXiv:0507227 [hep-th]].  
  
\bibitem{Andrianopoli:2007ep}
  L.~Andrianopoli, R.~D'Auria and L.~Sommovigo,
  ``D=4, N=2 supergravity in the presence of vector-tensor multiplets and the role of higher p-forms in the framework of free differential algebras,''
  Adv.\ Stud.\ Theor.\ Phys.\  {\bf 1} (2008) 561,
  [arXiv:0710.3107 [hep-th]].
  
\bibitem{Andrianopoli:2011zj}
  L.~Andrianopoli, R.~D'Auria, L.~Sommovigo and M.~Trigiante, 
  ``D=4, N=2 Gauged Supergravity coupled to Vector-Tensor Multiplets,'' 
  Nucl.\ Phys.\ B {\bf 851} (2011) 1, 
  doi:10.1016/j.nuclphysb.2011.05.007
  [arXiv:1103.4813 [hep-th]].
  
\bibitem{sito} 4d supergravity Lie 2-algebra in nLab, \\ https://ncatlab.org/nlab/show/4d+supergravity+Lie+2-algebra   
  

\bibitem{bacry} H.~Bacry, P.~Combe and J.~L.~Richard,
  ``Group-theoretical analysis of elementary particles in an external electromagnetic field. 1. the relativistic particle in a constant and uniform field,''
  Nuovo Cim.\ A {\bf 67} (1970) 267,
  doi:10.1007/BF02725178
  
\bibitem{schrader} R.~Schrader,
  ``The Maxwell group and the quantum theory of particles in classical homogeneous electromagnetic fields,''
  Fortsch.\ Phys.\  {\bf 20} (1972) 701, 
  doi:10.1002/prop.19720201202  
  
\bibitem{gomis0} J.~Gomis, K.~Kamimura and J.~Lukierski,
  ``Deformations of Maxwell algebra and their Dynamical Realizations,''
  JHEP {\bf 0908} (2009) 039, 
  doi:10.1088/1126-6708/2009/08/039
  [arXiv:0906.4464 [hep-th]].  
  
\bibitem{gomis1} J.~Gomis, K.~Kamimura and J.~Lukierski,
  ``Deformed Maxwell Algebras and their Realizations,''
  AIP Conf.\ Proc.\  {\bf 1196} (2009) 124, 
  doi:10.1063/1.3284373
  [arXiv:0910.0326 [hep-th]].   
  
\bibitem{Bonanos:2008ez}
  S.~Bonanos and J.~Gomis,
  ``Infinite Sequence of Poincare Group Extensions: Structure and Dynamics,''
  J.\ Phys.\ A {\bf 43} (2010) 015201,
  doi:10.1088/1751-8113/43/1/015201
  [arXiv:0812.4140 [hep-th]].  
  
\bibitem{gibbons}
  G.~W.~Gibbons, J.~Gomis and C.~N.~Pope,
  ``Deforming the Maxwell-Sim Algebra,''
  Phys.\ Rev.\ D {\bf 82} (2010) 065002,
  doi:10.1103/PhysRevD.82.065002
  [arXiv:0910.3220 [hep-th]].
  
\bibitem{deAzcarraga:2010sw}
  J.~A.~de Azcarraga, K.~Kamimura and J.~Lukierski,
  ``Generalized cosmological term from Maxwell symmetries,''
  Phys.\ Rev.\ D {\bf 83} (2011) 124036, 
  doi:10.1103/PhysRevD.83.124036
  [arXiv:1012.4402 [hep-th]].  
  
\bibitem{Durka:2011nf}
  R.~Durka, J.~Kowalski-Glikman and M.~Szczachor,
  ``Gauged AdS-Maxwell algebra and gravity,''
  Mod.\ Phys.\ Lett.\ A {\bf 26} (2011) 2689, 
  doi:10.1142/S0217732311037078
  [arXiv:1107.4728 [hep-th]].     
  
\bibitem{gomis2} S.~Bonanos, J.~Gomis, K.~Kamimura and J.~Lukierski,
  ``Maxwell Superalgebra and Superparticle in Constant Gauge Badkgrounds,'' 
  Phys.\ Rev.\ Lett.\  {\bf 104} (2010) 090401, 
  doi:10.1103/PhysRevLett.104.090401
  [arXiv:0911.5072 [hep-th]].     
  
\bibitem{gomis3} S.~Bonanos, J.~Gomis, K.~Kamimura and J.~Lukierski,
  ``Deformations of Maxwell Superalgebras and Their Applications,''  
  J.\ Math.\ Phys.\  {\bf 51} (2010) 102301, 
  doi:10.1063/1.3492928
  [arXiv:1005.3714 [hep-th]].     
  
\bibitem{Kamimura:2011mq}
  K.~Kamimura and J.~Lukierski,
  ``Supersymmetrization Schemes of D=4 Maxwell Algebra,''
  Phys.\ Lett.\ B {\bf 707} (2012) 292,
  doi:10.1016/j.physletb.2011.12.037
  [arXiv:1111.3598 [math-ph]].     
  
\bibitem{deAzcarraga:2012zv}
  J.~A.~de Azcarraga, J.~M.~Izquierdo, J.~Lukierski and M.~Woronowicz,
  ``Generalizations of Maxwell (super)algebras by the expansion method,'' 
  Nucl.\ Phys.\ B {\bf 869} (2013) 303, 
  doi:10.1016/j.nuclphysb.2012.12.008
  [arXiv:1210.1117 [hep-th]].  
  
\bibitem{Concha2} P.~K.~Concha and E.~K.~Rodr\'{i}guez,
  ``Maxwell Superalgebras and Abelian Semigroup Expansion,''
  Nucl.\ Phys.\ B {\bf 886} (2014) 1128, 
  doi:10.1016/j.nuclphysb.2014.07.022
  [arXiv:1405.1334 [hep-th]].
  
\bibitem{green} M.~B.~Green,
  ``Supertranslations, Superstrings and Chern-Simons Forms,''
  Phys.\ Lett.\ B {\bf 223} (1989) 157,
  doi:10.1016/0370-2693(89)90233-5
  
\bibitem{Siegel:1994xr}
  W.~Siegel,
  ``Randomizing the superstring,'' 	
  Phys.\ Rev.\ D {\bf 50} (1994) 2799, 		
  doi:10.1103/PhysRevD.50.2799
  [arXiv:9403144 [hep-th]].  
  
\bibitem{Bergshoeff:1995hm}
  E.~Bergshoeff and E.~Sezgin,
  ``Superp-Brane theories and new space-time superalgebras,''
  Phys.\ Lett.\ B {\bf 354} (1995) 256,
  doi:10.1016/0370-2693(95)00655-5
  [arXiv:9504140 [hep-th]].
  
\bibitem{Sezgin:1996cj}
  E.~Sezgin,
  ``The M algebra,''
  Phys.\ Lett.\ B {\bf 392} (1997) 323,
  doi:10.1016/S0370-2693(96)01576-6
  [arXiv:9609086 [hep-th]].    
  
\bibitem{deAzcarraga:2014jpa}
  J.~A.~de Azcarraga and J.~M.~Izquierdo,
  ``Minimal D = 4 supergravity from the superMaxwell algebra,''
  Nucl.\ Phys.\ B {\bf 885} (2014) 34,
  doi:10.1016/j.nuclphysb.2014.05.007
  [arXiv:1403.4128 [hep-th]].

\bibitem{CR2} P.~K.~Concha and E.~K.~Rodr\'{i}guez,
  ``N = 1 Supergravity and Maxwell superalgebras,''
  JHEP {\bf 1409} (2014) 090, 
  doi:10.1007/JHEP09(2014)090
  [arXiv:1407.4635 [hep-th]].
  
\bibitem{Concha:2016hbt} P.~K.~Concha, R.~Durka, N.~Merino and E.~K.~Rodr\'{i}guez,
  ``New family of Maxwell like algebras,''
  Phys.\ Lett.\ B {\bf 759} (2016) 507, 
  doi:10.1016/j.physletb.2016.06.016
  [arXiv:1601.06443 [hep-th]].  
  
\bibitem{Castellani} L.~Castellani, P.~Fr\'{e}, F.~Giani, K.~Pilch and P.~van Nieuwenhuizen,
  ``Gauging of $d=11$ Supergravity?,''
  Annals Phys.\  {\bf 146} (1983) 35, 
  doi:10.1016/0003-4916(83)90052-0  

  
\bibitem{deAzcarraga:1989mza}
  J.~A.~de Azcarraga, J.~P.~Gauntlett, J.~M.~Izquierdo and P.~K.~Townsend,
  ``Topological Extensions of the Supersymmetry Algebra for Extended Objects,''
  Phys.\ Rev.\ Lett.\  {\bf 63} (1989) 2443,
  doi:10.1103/PhysRevLett.63.2443

\bibitem{Townsend:1997wg}
  P.~K.~Townsend,
  ``M theory from its superalgebra,''
  In *Cargese 1997, Strings, branes and dualities* 141-177,
  [arXiv:9712004 [hep-th]].

\bibitem{Hassaine:2003vq}
  M.~Hassaine, R.~Troncoso and J.~Zanelli,
  ``Poincare invariant gravity with local supersymmetry as a gauge theory for the M-algebra,''
  Phys.\ Lett.\ B {\bf 596} (2004) 132,
  doi:10.1016/j.physletb.2004.06.067
  [arXiv:0306258 [hep-th]].

\bibitem{Hassaine:2004pp}
  M.~Hassaine, R.~Troncoso and J.~Zanelli,
  ``11D supergravity as a gauge theory for the M-algebra,''
  PoS WC {\bf 2004} (2005) 006,
  [arXiv:0503220 [hep-th]].  
 

\bibitem{Gomis:2017cmt}
  J.~Gomis and A.~Kleinschmidt,
  ``On free Lie algebras and particles in electro-magnetic fields,''
  JHEP {\bf 1707} (2017) 085,
  doi:10.1007/JHEP07(2017)085
  [arXiv:1705.05854 [hep-th]].

\bibitem{Durka:2011gm}
  R.~Durka, J.~Kowalski-Glikman and M.~Szczachor,
  ``AdS-Maxwell superalgebra and supergravity,''
  Mod.\ Phys.\ Lett.\ A {\bf 27} (2012) 1250023,
  doi:10.1142/S021773231250023X
  [arXiv:1107.5731 [hep-th]].    

\end{thebibliography}
\end{document}